\documentclass[10pt]{article}
\usepackage{amsmath}
\usepackage{amssymb}
\usepackage{latexsym}
\usepackage{wasysym}
\usepackage{longtable}

\begin{document}
\title{Coordinate Data and Identification of Stars in Astrophysical Catalogs}
\author{Nuriya T. Ashimbaeva, Sternberg Astronomical Institute}
\maketitle
\begin{abstract}

The aim of our paper is to compile high-precision positional star catalogs by combining extensive astrophysical data sets with the results of large astrometric surveys. The essential features required of the compiled catalogs are data reliability and unambiguity. There are cases where the commonly adopted accuracy of 1 arc sec in coordinates cannot ensure unambiguous identification of stars. Therefore methods of verification of catalogs have become increasingly important, which require a certain technique of unambiguous identification of objects. This paper reports the results of identification of stars in The Henry Draper Extension Charts catalog and in the catalogs of variable stars using the identification technique developed. Examples of the resolution of conflicts involving double and multiple stars in astrophysical catalogs are discussed.
\end{abstract}

\section*{Introduction.}

The ever increasing amount of the data from various fields of 
both astrophysics and Galactic astronomy on the one hand, and 
the release of astrometric catalogs containing accurate coordinates 
for millions of stars (HIPPARCOS, TYCHO, USNO, UCAC, and 2MASS), 
on the other hand, provide good prerequisites for the creation 
of compiled catalogs for various purposes containing high-precision 
positions of the objects studied. 

The principal requirement toward the catalogs compiled are reliability 
and unambiguity of the data. The ever increasing sizes of the 
catalogs impose increasingly stringent criteria. The currently 
generally adopted accuracy ($1''$ or better) has become insufficient, 
because it fails to guarantee unambiguous identification of objects, 
e.g., in the cases of close optical binaries or star clusters.

No data can be used by itself without invoking adequate processing 
algorithms, and therefore the development of a technique for 
reliable and unambiguous identification of objects has become 
a task of great importance and urgency. The form of the initial 
data prevented the use of automated identification procedures 
(Nesterov \textit{et al}, 1995, Gulyaev and Ashimbaeva, 1997), making 
it necessary to perform manual identification of each object.

\section{Technique of identification of stars in the HDEC catalog 
in the catalogs of variable stars.}

In the process of our work we developed a set of rules for identification 
of objects. Like in the case of automatic identification we set 
a certain interval of permissible values. 

The object usually must meet a number of criteria simultaneously:

\begin{enumerate}

\item We set certain coordinate tolerances (the minimal ``box'', 
where identification is possible without loss of information; 
maximum size of the field studied whose increase would prevent 
the identification of the configuration);

\item %(2) 
Photometric indices (we set a magnitude interval); 

\item %(3) 
Allowance for the proper motions;

\item %(4) 
Referencing the object to star configurations (the configuration 
of the identified object must strictly agree with that depicted 
in the published figure).

\end{enumerate}

Too stringent requirements increase the risk of losing information 
about the object. Too loose requirements, on the other hand, 
increase the risk of getting too much unnecessary and erroneous 
information. Hence we need to optimize the criteria imposed. 
The final identification must be unambiguous even for very approximate 
input data (with coordinates known with errors of up to 1 arcmin 
or even more).

\section{Determination of the coordinates of the stars of the 
Second Extension of the HD catalog.}

The work on accurate identification of astronomical objects by 
assigning to them accurate coordinates adopted from mass astrometric 
catalogs began in 1990-ies. The results of the determination 
of the coordinates for HDEC stars can be found in (Nesterov \textit{et 
al}. 1995). Here is a brief review of the subject history. The 
second extension of the HD catalog (Cannon A.J., Mayall M. W., 
1949) was published in the form of 275 charts, and has therefore 
been almost totally neglected by the international research community. 
A number of researches attempted to convert the Second extension 
of the HD into a more convenient form: Bonnet (1978), Roman (1992), R\"oser et al. (1991). The latter authors measured the Cartesian 
coordinates for a total of 10 639 stars of the  HDEC. The average 
position accuracy of these measurements was $0''.3$. This 
work demonstrated that it was possible in principle to convert HDEC 
data into a computer-readable form.

We identified each star manually, because automatic reduction 
was evidently out of question due to the form of the initial 
data. We identified stars by comparing the published chart to 
the image drawn by SimFOV program (A.A.Volchkov and O.A.Volchkov), 
which visualizes the positions of stars from the reference catalog 
on a computer monitor. We used the  GSC-AC catalog (\textit{R\"oser 
et al, 1995; Nesterov and Gulyaev, 1992}) as our reference catalog. 
We assigned to each HDE star the best-matching GSC-AC star according 
to visual inspection.

We then compared the initial data set with the measurements of 
the Cartesian coordinates of HDEC stars made at Astronomische 
Recheninstitut (Heidelberg, Germany). As a result of additional 
comparison and checks we obtained a data set of very high quality.

In addition to single stars we also found 1783 multiple systems 
(or blends). A total of 503 variable stars make up a separate 
list. 

\textbf{Result}. The HDEC star catalog is available from Strasbourg 
Data Center as catalog III/182. The result of the work on the 
HDEC catalog can be summarized as follows:

(1) The second HD extension, which hitherto existed in the form 
of finding charts exclusively has been converted into a star 
catalog.

(2) We determined precise (to within $0.5''$) positions for 86 
933 HDEC stars; proper motions accurate to about 5 milliarcseconds/yr 
are given for more than 96 \% of all stars.

\section{Creation of the astrometric catalog of variable stars.}

Our next step was to determine the accurate coordinates for variable 
stars using the identification technique described above. Since 
1994, we started compiling our catalog of positions and proper 
motions (Gulyaev and Ashimbaeva, 1997). 

Our main source of data on variable stars was the Fourth edition 
of the General Catalog of Variable Stars (Kholopov et al.,1985-1988). 
The object coordinates given in this catalog were formally accurate 
to within $1^{s}$ and $0^{m}.1$ in $\alpha$ and $\delta$, respectively, 
although the actual situation proved to be much worse.

Again, stars of the General Catalog of Variable Stars (GCVS) 
were impossible to identify automatically due to the form of 
the data presented --- only finding charts of different quality 
were available. We therefore identified stars using the technique 
already tested with HDEC. We compared each finding chart with 
the corresponding stellar field drawn on the computer monitor 
based on the data from the available reference catalog.

We cross identified variable stars with: (1) the PPM star catalog 
(R\"oser et al., 1993) for the brightest stars and (2) the first 
release of the GSC/AC catalog (Nesterov and Gulyaev, 1992), for 
the bulk of the stars.

\textbf{Results.} Table 1 summarizes the quantitative results 
of identification for the catalog considered.

\begin{table}	%Table 1.
\caption{Results of identification.}
{\small
\begin{tabular}{|l|r|}
\hline
% ROW 1
{\raggedright Total number of stars in the GCVS including name lists nos. 67-72} & 
{\raggedleft 31 081} \\
\hline
% ROW 2
{\raggedright - stars with available finding charts} & 
{\raggedleft 26 797} \\
\hline
% ROW 3
{\raggedright Number of stars in the NSV catalog with identified finding charts} & 
{\raggedleft 7 991} \\
\hline
% ROW 4
{\raggedright Total number of finding charts examined} & 
{\raggedleft 42 789} \\
\hline
% ROW 5
{\raggedright Number of stars identified} & 
{\raggedleft 21 789} \\
\hline
% ROW 6
{\raggedright Number of missing stars with identified finding charts} & 
{\raggedleft 12 257} \\
\hline
% ROW 7
{\raggedright Number of GCVS stars whose finding charts could not be identified.} & 
{\raggedleft 742} \\
\hline
\end{tabular}
}
\end{table}

The catalog compiled by 1997 had the form of a data set containing 
identifications with the PPM, GSC, and AC catalogs. Because of 
the death of A.P.Gulyaev in 1998 the work remained incompleted 
with no electronic version has been produced. We later reconstructed 
the catalog based on the available identification files and numerous 
archive records (Ashimbaeva et al., 2004). In addition to the 
previously employed GSC and AC=4M catalogs we identified variable 
stars with the HIPPARCOS (The Hipparcos and Tycho Catalogues, 
1997), TYCHO2 (Hog et al, 2000), NPM (Klemola et al.,1987),
and A2.0 (Monet et al., 1998) catalogs. We greatly appreciate 
the invaluable assistance and advice offered by the members of 
the Division of Variable Stars of the Sternberg Astronomical 
Institute and especially by N.N.Samus.

The catalog is available at {\underline{http://astrometric.sai.msu.ru/lib\_projects\_01.html}}. 
Here are a few comments concerning its contents. 

Stars in the catalog are sorted by their names in accordance 
with the GCVS convention. NSV stars (Kukarkin et al. 1982) are 
sorted by their numbers and listed after the GCVS stars. The 
accuracy of coordinates depends on the particular reference catalog 
employed and varies from $0.5''$ for GSC to $0.0006''$ for HIPPARCOS. 
The J2000.0 equatorial coordinates are given with the following 
accuracy: right ascensions --- to within 0$^{s}$.001 and declinations, 
to within $0''.01$. The proper motions are given as in the 
corresponding source catalog, i.e., to $0''.00001$/yr for 
the HIPPARCOS catalog and to $0''.001$/yr for the Astrographic 
catalog. The coordinate epoch is given in the cases where no 
proper-motion data are available. For the convenience of identification 
we also give the number of the star according to the HIPPARCOS, 
NPM, Tycho2, or GSC catalog. We also give the photometric data 
(magnitude, magnitude interval) as provided by the reference 
catalogs used for identification (HIPPARCOS, NPM, Tycho2, GSC, 
AC, PPM, and A2.0).

\textbf{Results of the work.} We refined the coordinates for a total 
of 21 971 variable stars with the positional and proper-motion 
accuracy matching those of the astromeric reference catalogs.

\section{Multiple systems in the HDEC.}

We analyzed multiple stars among the HDEC objects. To this end, 
we used Aladin program of the Astronomical Data Center in Strasbourg 
(http://aladin.u-strasbg.fr/).

Our analysis allowed us to: (1) refine the coordinates of each 
component in accordance with the data provided by the reference 
catalogs and measure the coordinates on the image; (2) identify 
all false measurements -- in the cases where the star proved 
to be single or wrong coordinates were given for the component; 
(3) determine the proper motions for a number of stars by analyzing 
the images taken at different epochs, and (4) sometimes find 
new components by comparing charts made at different wavelengths 
and for different epochs. The reduction of the entire data set 
yielded more than 4000 components. In addition to component identification 
errors we also revealed gross errors of the HDEC, which are inevitable 
in a large catalog.

\textit{We compared our catalog to other CDS catalogs in order 
to search for the available identifications for multiple stars 
of HDEC. Table 2 summarizes the results of this comparison. }

\begin{flushright}

\textbf{Table} \textbf{2.} Comparison of identification of multiple stars 
of the HDEC catalog.\\
\end{flushright}

\begin{longtable}{|p{1.975in}|p{2.525in}|}
\hline
% ROW 1
{\centering \textbf{Catalog}} & 
{\centering \textbf{Identifications}} \\
\hline
% ROW 2
{\raggedright HIPPARCOS input catalog} & 
{\raggedright 23 multiple stars of the HDEC catalog} \\
\hline
% ROW 3
{\raggedright HIPPARCOS and TYCHO catalogs (they are now inseparable)} & 
{\raggedright 822 multiple stars of the HDEC catalog} \\
\hline
% ROW 4
{\raggedright TYCHO-2} & 
{\raggedright 1007 identifications by numbers, 14 stars of the catalog have 
CCDM component designations} \\
\hline
% ROW 5
{\raggedright TYCHO-3} & 
{\raggedright 350 identifications with different numbers including 21 double 
identifications.} \\
\hline
% ROW 6
{\raggedright Visual Double Stars in HIPPARCOS (Dommanget et al., 2000)} & 
{\raggedright 17 multiple systems and 54 components} \\
\hline
% ROW 7
{\raggedright The Tycho-2 Spectral Type Catalog} & 
{\raggedright Identification by coordinates ($2''$ identification radius)? 
135 objects. \linebreak
Identification by number (5\ensuremath{''} identification radius)- 102 
stars.} \\
\hline
\end{longtable}

We found a total of 319 single stars including 11 stars with 
proper motions; assigned coordinates to all components with an 
accuracy of $0.''1$; assigned spectra to individual components 
in 1166 cases; found 611 double and eight triple and multiple 
blends; revealed errors in HDEC data records; found identification 
errors in the system of the Astrographic catalog, and compared 
our data with those of other catalogs of multiple systems.

\section{Conclusions.}

The aim of this work was to produce compiled catalogs containing 
high-precision coordinates of interesting objects, which have 
other data published in astrophysical catalogs such as HD and 
General Catalog of Variable Stars. As a result, we determined 
for the first time accurate coordinates and proper motions for 
objects of the Second Extension of HD and refined coordinates 
and proper motions for objects of the General Catalog of Variable 
Stars.

We developed a technique for nonautomatic identification of objects, 
which facilitates and speeds up the process. We continued our 
work by analyzing complex cases of identification of multiple 
systems in the HDEC catalog. We found cases of false identification, 
revealed possible causes of such errors, and refined the coordinates 
of individual components.

\section*{References.}

\textit{Ashimbaeva} \textit{N.T., Volchkov A.A, Romanova G.V.} Astrometric 
Catalog of Variable Stars.// Trudy Sternberg Astron. Inst.. 2004. 
V.70. P.287. 

\textit{Volchkov} \textit{A.A., Volchkov O.A}. SimFOV //{\underline{http://www.simfov.ru/description/}}.

\textit{Gulyaev} \textit{A.P., Ashimbaeva N.T.} Astromeric Catalog of Variable 
Stars // Astron. Zh., 1997, V.74, No. 2, P.249-253.

\textit{Kukarkin} \textit{B.V., Kholopov P.N., Artyukhina N.M.} // Catalog 
of Suspected Variable Stars. Moscow: Nauka. 1982.

\textit{Nesterov} \textit{V.V.. Gulyaev A.P.} // O Chetyrekhmillionnom 
Kataloge Zvezd (The Four-Million Star Catalog). Moscow: Moscow 
State University, 1992.

\textit{Kholopov} \textit{P.N., Samus' N.N., Goranskii V.P, et al..} // 
General Catalogue of Variable Stars, Fourth edition. Moscow: 
Nauka. 1985-1988. Vols.I-III.

\textit{Bastian U., R?eser S., Yagudin L., Nesterov V. PPM Star Catalogue: 
Positions and Proper Motions of 197,179 stars south of -2.5\ensuremath{^\circ} 
declination for equinox and epoch J2000.0. Vol.III-IV. Heidelberg; 
Berlin; Oxford: Spektrum, Acad. Verl., 1993.}

\textit{Cannon A.J., Mayall M. W. //} Annie J. Cannon Memorial Volume 
of the Henry Draper Extension. Ann. Astron. Obs. Harv. Coll. 
1949. V. 112.

\textit{Carney B.W., Latham D.W., Laird J.B., Aguilar L.A.} Proper 
motion stars survey. XII. //Astron. J. 1994. V.107. P. 2240.

\textit{Giclas H.L., Burnham Jr. R., Thomas N.G.} Lowell Proper Motion 
Survey: 8991 Stars with $m>8$, $PM>0.26''$/year 
in the Northern Hemisphere. //Lowell Obs., Flagstaff, AZ. 1971.

\textit{Giclas H.L., Burnham Jr. R., Thomas N.G.} Lowell Proper Motion 
Survey - Southern Hemisphere. //Lowell Observatory Bulletin. 
1978. No. 164.

\textit{Hog E., Fabricius C., Makarov V.V., Urban S., Corbin T., 
Wycoff G., Bastian U., Schwekendiek P., Wicenec A.} The Tycho-2 
Catalogue of the 2.5 Million Brightest Stars. // Astron. Astrophys. 
2000. V. 355, L27.

\textit{Klemola A. R., Hanson R. B., and Jones B. F.,} The Lick Northern 
Proper Motion Program // Centre de Donne?es Astronomiques de 
Strasbourg, 1987. I/199.

\textit{Monet D., Bird A., Canzian B., et al.,} USNO-A V2.0. A Catalog 
of Astrometric Standards. 1998. US Naval Observatory,Washington.

\textit{Nesterov V.V., Kuzmin A.V., Ashimbaeva N.T., Volchkov A.A., 
Roeser S., Bastian U.} The Henry Draper Extension Charts: A catalog 
of accurate positions, proper motions, magnitudes and spectral 
types of 86933 stars. // Astron. Astrophys. Suppl. Ser. 1995. 
V.110. P.367.

\textit{R\"oser S., Bastian U. PPM Star Catalogue: Positions and Proper 
Motions of 181,731 stars north of $-2.5^\circ$ declination 
for equinox and epoch J2000.0. Vol. I-II. Heidelberg; Berlin; 
Oxford: Spektrum, Acad. Verl., 1991.}

\textit{R\"oser, S., Bastian, U., and Kuzmin, A. V.} I.A.U. Colloquium 
148, ASP Conf. Series, 1995, Vol 84, ed. J. M. Chapman et al.

\textit{R\"oser S., Bastian U., Wiese K.} The Second Henry Draper Extension: 
Spectral Types and Precise Positions for 10639 Stars. //Astron. 
Astrophys. Suppl. Ser. 1991. V.88. P. 277.

The Hipparcos and Tycho Catalogues (European Space Agency, 1997), 
ESA SP-1200.

\end{document}